\documentclass[
    aip, jap,
    reprint,
    english,
    superscriptaddress,
    longbibliography,
]{revtex4-1}

\usepackage[utf8]{inputenc}
\usepackage[T1]{fontenc}

\usepackage{graphicx}
\graphicspath{{Figs/}}
\usepackage{hyperref}

\usepackage{
    amsmath,
    amssymb,
    gensymb,
}

\begin{document}



    \title{
        Temperature dependence of exchange biased multiferroic \( \mathbf{BiFeO_3} \)\slash \( \mathbf{Ni_{81}Fe_{19}} \) polycrystalline bilayer
    }

    \author{J. Richy}
    \affiliation{
        Laboratoire de Magnétisme de Bretagne, Department of Physics, CNRS-UBO, 29285 Brest Cedex 3, France
    }
    \affiliation{
        Department of Physics, University of Johannesburg, PO Box 524, Auckland Park, Johannesburg 2006, South Africa
    }

    \author{T. Hauguel}
    \author{J. Ph. Jay}
    \author{S. P. Pogossian}
    \affiliation{
        Laboratoire de Magnétisme de Bretagne, Department of Physics, CNRS-UBO, 29285 Brest Cedex 3, France
    }

    \author{B. Warot-Fonrose}
    \affiliation{
        CEMES CNRS-UPR 8011, Université de Toulouse, 31055 Toulouse, France
    }

    \author{C. J. Sheppard}
    \author{J. L. Snyman}
    \author{A. M. Strydom}
    \affiliation{
        Department of Physics, University of Johannesburg, PO Box 524, Auckland Park, Johannesburg 2006, South Africa
    }

    \author{J. Ben~Youssef}
    \affiliation{
        Laboratoire de Magnétisme de Bretagne, Department of Physics, CNRS-UBO, 29285 Brest Cedex 3, France
    }

    \author{A. R. E. Prinsloo}
    \affiliation{
        Department of Physics, University of Johannesburg, PO Box 524, Auckland Park, Johannesburg 2006, South Africa
    }

    \author{D. Spenato}
    \author{D. T. Dekadjevi}
    \email[]{david.dekadjevi@univ-brest.fr}
    \affiliation{
        Laboratoire de Magnétisme de Bretagne, Department of Physics, CNRS-UBO, 29285 Brest Cedex 3, France
    }

    \date{\today}

    \begin{abstract}
        The temperature dependence of exchange bias properties are studied in polycrystalline \bfo/\py bilayers, for different \bfo thicknesses.
        Using a field cooling protocol, a non-monotonic behavior of the exchange bias field is shown in the exchange-biased bilayers.
        Another thermal protocol, the Soeya protocol, related to the \bfo thermal activation energies was carried out and reveals a two-step evolution of the exchange bias field.
        The results of these two different protocols are similar to the ones obtained for measurements previously reported on epitaxial \bfo, indicating a driving mechanism independent of the long-range crystalline arrangement (i.e., epitaxial or polycrystalline).
        An intrinsic property of \bfo is proposed as being the driving mechanism for the thermal dependent magnetization reversal: the canting of the \bfo spins leading to a biquadratic contribution to the exchange coupling.
        The temperature dependence of the magnetization reversal angular behavior agrees with the presence of such a biquadratic contribution for exchange biased bilayers studied here.
    \end{abstract}

    \pacs{
    }

    \maketitle


\section{Introduction}

    Electrical control of magnetic nanostructures would create a new generation of electronic devices directly integrable in actual device architecture.%
    \cite{Matsukura-2015-ID563}
    A great deal of research has been focused on an efficient way to control magnetic properties using an electric field, with no need of an applied magnetic field.%
    \cite{Chiba-2008-ID558, Wu-2010-ID562, Heron-2011-ID333}
    This research is of importance in the field of applied physics when considering magnetic memories or high frequency devices.%
    \cite{Chu-2008-ID109, Eerenstein-2006-ID108, Bibes-2008-ID110}
    For example, spin polarized current is an effective mechanism to transfer a torque to magnetization.
    However, it requires large current densities, leading to energy loss because of Joule heating effects.%
    \cite{Matsukura-2015-ID563}
    Among the different possibilities of utilizing an electric control of magnetic properties, the use of single-phase magnetoelectric multiferroics (MMF) is considered, as these allow a direct means of controlling magnetization via an electric field in a single heterostructure.%
    \cite{Bibes-2008-ID110, Zhao-2006-ID160, Roy-2012-ID177, Vaz-2012-ID393}
    Room temperature MMF materials are rare.
    Among these, \bfo (BFO) is one of the most studied because of its high ferroelectric (FE) polarization, with a FE Curie temperature in the order of \SI{1100}{K}.%
    \cite{Bea-2007-ID82, Catalan-2009-ID134}
    In addition, the BFO possesses an antiferromagnetic (AF) order with a Néel temperature of about \SI{640}{K}.%
    \cite{Connolly-1972-ID569}
    In order to use an AF magnetoelectric material with no net magnetization, a ferromagnetic/AF exchange coupling is proposed.
    Meiklejohn and Bean\cite{Meiklejohn-1956-ID35} found it can be introduced by placing the AF layer in contact with the ferromagnetic (F) material, and is refered to as exchange bias coupling.
    It produces an additional anisotropy that stabilize the F layer.
    The existence of exchange bias coupling is revealed by a field shift \He of the hysteresis cycle, named \emph{exchange bias} field, and by a coercive enhancement.
    In exchange biased (EB) systems, \He originates from the interfacial pinned spins, that is, the non-reversible part of interfacial spins.
    The \Hc enhancement originates in the reversible process driven by the AF anisotropy.
    This anisotropy provides additional critical fields that will hinder the domain wall motion in the F layer.%
    \cite{Misra-2004-ID65, Li-2000-ID619}
    In recent years many research projects have been undertaken to understand the exchange bias coupling in BFO/F nanostructures.%
	\cite{Ederer-2005-ID135, Chu-2007-ID190, Chu-2009-ID560, Heron-2014-ID228, Lebeugle-2010-ID8}
    However, there is still no clear understanding of the origins of this coupling.

    Among the properties of interest, the thermal dependence of magnetization reversal is a key phenomenon that need to be understood in AF/F systems due to its relevance for applied issues in magnetic recording, and for fundamental issues related to the thermodynamics of nanoscale magnetic systems.
    Effectively, understanding and tailoring the thermal dependence of the magnetization reversal is of interest for applied issues, as this dependence can be inferred by laser heating%
    \cite{Katayama-2000-ID620}
    or applied current.%
    \cite{Zhang-2002-ID343}
    In addition, it should be considered that a device can be compromised by temperature fluctuations in a magnetic field.%
    \cite{Weller-1999-ID488}
    Consequently, considering the particular interest of BFO/F systems, the thermal dependence of the BFO/F magnetization reversal needs to be understood.
    Previous experimental studies on BFO/F systems have revealed an intriguing non-monotonic evolution of the exchange bias field with temperature \T.%
    \cite{Martin-2008-ID112, Naganuma-2011-ID127, He-2015-ID326, Xue-2013-ID83}
    Furthermore, this phenomenon is found in epitaxial%
    \cite{Martin-2008-ID112, Naganuma-2011-ID127, He-2015-ID326}
    and polycrystalline%
    \cite{Xue-2013-ID83}
    bilayers, for different F (including CoFe, CoFeB, Co and NiFe) coupled with BFO, and for different BFO thicknesses.
    This common phenomenon in BFO/F systems is not yet understood despite its key importance.

    In the research work presented here, the thermal behavior of polycrystalline \bfo\slash \py is studied for different BFO thicknesses (\tbfo), using complementary approaches.
    Results are analyzed considering previous findings on the thermal magnetization reversals of not only polycrystalline but also epitaxial BFO/F systems, so as to gain a better understanding of the origin of the non-monotonic evolution of the exchange bias field with temperature.

    In the first section of this manuscript experimental procedures are described.
    In the second section, a structural study analysis involving transmission electron microscopy (TEM) and two dimensional (2D) temperature dependent X-ray diffraction (XRD) is provided.
    This was used to probe the crystallographic and the morphologic properties of the BFO/F bilayers.
    In the third section, the thermal dependence of magnetization reversals, including exchange bias and coercive fields evolutions, are studied using a field-cooled protocol in order to probe the presence of a non-monotonic behavior in the samples.
    In the fourth section, the exchange bias field evolution following a specific protocol, called here the Soeya protocol,\cite{Soeya-1994-ID14} is showed.
    This protocol, initially proposed by Soeya~\textit{et al.},\cite{Soeya-1994-ID14} and later modified by O'Grady~\textit{et al.},\cite{OGrady-2010-ID87} was used to probe the thermal activated energies of all BFO/F samples in this study.
    It should be noted that this paper is the first study to investigate this thermal dependent protocol in polycrystalline exchange-biased BFO/F samples.
    Finally, the thermal dependence of the magnetization reversal angular evolution is probed, as this will provide information on the various effective anisotropies of exchange-bias systems.

\section{Experimental procedures}

    The  heterostructures \bfo/\py (BFO/Py) were grown by radio frequency sputtering, sequentially deposited using the structure Pt(\SI{30}{nm})/\allowbreak BFO(\( t_\mathrm{BFO} \))/\allowbreak Py(\SI{10}{nm})/\allowbreak Pt(\SI{30}{nm}) on a naturally oxidized silicon substrate.
    An in-plane deposition field \( \Hdep = \SI{30}{mT} \) magnetic field was applied during the growth process.
    Further details on the growth process are available in previous publications.%
    \cite{Hauguel-2011-ID4, Hauguel-2012-ID11}
    The BFO nominal thicknesses were equal to \SI{0}{nm}, \SI{29}{nm} and \SI{177}{nm}, further referred to as sample \Sa , \Sb and \Sc.
    The BFO critical thickness \tc above which \He is not zero, was determined earlier to be \SI{23}{nm} in the BFO/Py system presented here.%
    \cite{Hauguel-2012-ID11}
    Thus, these samples are representative of three different thickness intervals of \( \He(\tbfo) \):
    i) \Sa corresponds to an unbiased sample;
    ii) \Sb with \( \tbfo = \SI{29}{nm} \) is just above \tc, an interval where \( \He(\tbfo) \) is strongly thickness dependent;
    and iii) \tbfo of \Sc with \( \tbfo = \SI{177}{nm} \) is far larger than \tc, an interval where \( \He(\tbfo) \) is thickness independent.

    In order to characterize the samples' structural properties, TEM and XRD measurements were carried out.
    TEM analyses on the samples were done using a TECNAI F-20 system operating at 200~kV, equipped with a spherical aberration corrector for the objective lens in order to avoid the delocalization effect at interfaces.
    The crystallographic properties of the samples were probed by XRD, using a 2D Oxford diffractometer (X-Calibur-2 model) in the temperature range from 100~K to 300~K.

    Temperature dependent magnetic measurements were initially performed using a superconductive quantum interference device (SQUID) magnetometry.
    In order to correct the data for remnant fields that might exist in the SQUID magnet, care was taken during the measurement protocols to correct for this.
    Two different measurement protocols were performed using the SQUID.
    These two protocols were i) the field cooled (FC) protocol, and ii) the Soeya protocol,\cite{Soeya-1994-ID14} that will be discussed in detail later in the text.

    In the FC protocol, the samples were cooled from 300~K to 10~K, in a \( \mu_0\Fc = \SI{100}{mT} \) field along the \Hdep direction.
    In order to dispose of all possible training effects, four field-switchings were performed at 10~K between negative and positive \Fc.
    The magnetic hysteresis (M-H) loops were then recorded at different temperatures between 10~K to 380~K, in increasing temperature steps.

    In the Soeya protocol, the samples were heated under a positive magnetic field \Fc applied along \Hdep at a preset temperature \( \Ts = \SI{380}{K} \), and subsequently cooled under the same \Fc down to a measurement temperature \( \Tm = \SI{10}{K} \).
    Then, after reversing the field to \( -\Fc \), the system was annealed at an intermediate activation temperature \Ta, leading to the reversal of AF entities with low energy compared to the thermal energy at \Ta.
    Finally, returning to \Tm, an M-H loop was measured.
    This process was repeated for different \Ta values.
    In the Soeya protocol, \( \He(\Ta) \) depends on the thermal activation energies present in the magnetic system.
    Indeed, above a given temperature defined as a blocking temperature, an energy barrier in the AF is overcome by the thermal agitation.
    In a AF/F system, different kinds of energy barriers can be expected to exist, such as: i) anisotropy energy barriers related to grain sizes or magnetic domain sizes, ii) domain wall nucleation and depinning energies, and iii) magnetic coupling energies (including complex interfacial couplings such as a spin-glass like coupling).
    Consequently, a distribution of the blocking temperatures can arise as a consequence of different phenomena involved.
    In principle, the evolution of \( \He(\Ta) \) may be attributed to any of the fore-mentioned energy barriers in EB systems.%
	\cite{Baltz-2010-ID126, Safeer-2012-ID86, OGrady-2010-ID87, Ventura-2007-ID605}

    In order to probe the temperature dependent anisotropy configuration, M-H loops were measured for different applied field directions \( \varphi \), where \( \varphi \) is the azimuthal angle with respect to \Hdep, using a custom made vibrating sample magnetometer (VSM).
    These measurements were done at room temperature (RT) and 77~K.
    In order to perform the 77~K measurements, the samples were immerged in liquid nitrogen.

\section{Results and Discussions}
\subsection{Structural properties}

    \begin{figure}
        \includegraphics{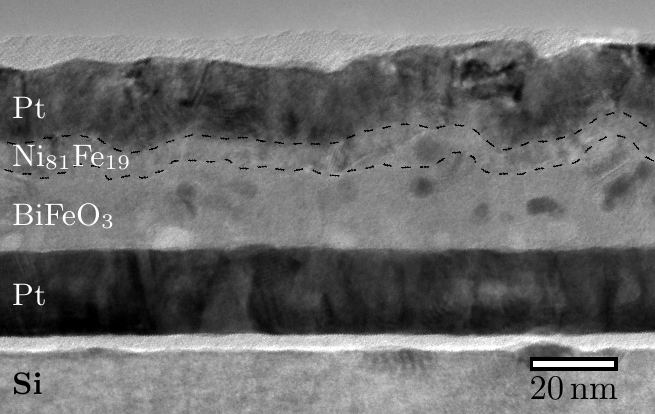}
        \caption{
            Cross-section transmission electron micrography of the \Sb sample.
            Dashed lines are guides to the eye, and delimit the Py layer.
        }
        \label{fig:1}
    \end{figure}

    In Fig.~\ref{fig:1} the TEM image of the \Sb sample cross-section is shown.
    It can be seen from Fig.~\ref{fig:1} that the various layers are well defined within the film layer stack.
    The surface roughness of the Py layer is due to the surface roughness of the BFO.
    TEM analysis of the \Sc sample shows an increase in the roughness of the Py layer when compared to that observed in \Sb, confirming an increase in the surface roughness with the increase of \tbfo, previously observed using atomic force microscopy measurements for \( \tbfo > \SI{23}{nm} \) on this BFO/Py system.%
    \cite{Hauguel-2012-ID11}

\begin{figure}
        \includegraphics{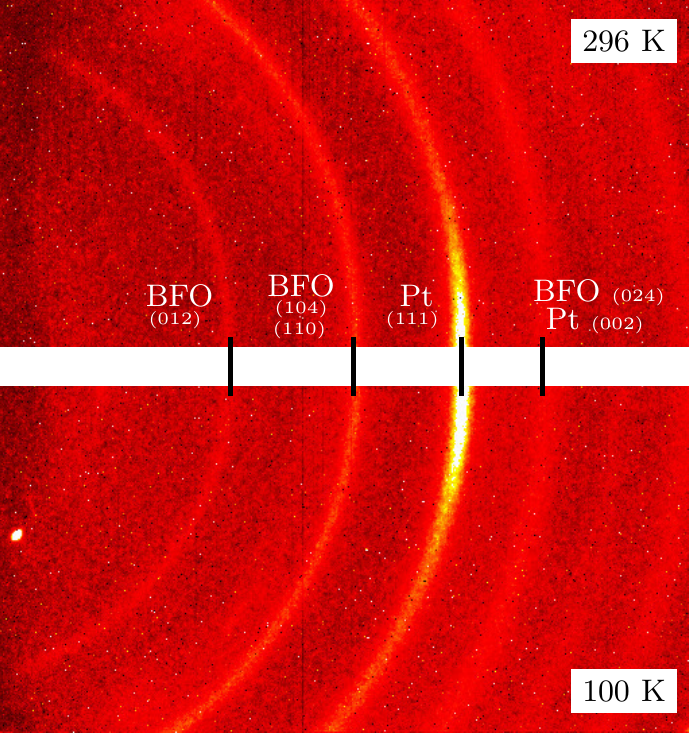}
        \caption{
            XRD patterns at two different temperatures for the \Sc heterostructure.
            The circular rings show the BFO polycrystallinity, and the non-uniform (111) Pt diffraction ring reveals a preferred growth direction for the Pt layer.
        }
        \label{fig:2}
    \end{figure}

	In order to investigate the BFO crystalline arrangement and its evolution with temperature, 2D XRD patterns were obtained on sample \Sc at 50~K intervals in the temperature range from 100~K to 296~K.
    Examples of these XRD patterns are shown in Fig.~\ref{fig:2}.
    It revealed circular rings typical of polycrystalline layers.
    The XRD patterns at 296~K and 100~K indicate a non-homogeneous intensity of the rings for the Pt layer.
    This is due to the preferred (111) growth direction of the Pt layer.
    There was no temperature dependence in these XRD diagrams, indicating the temperature stability of the BFO crystallographic arrangement.

\subsection{Temperature dependent magnetization reversal using the FC protocol}

\begin{figure}
        \includegraphics{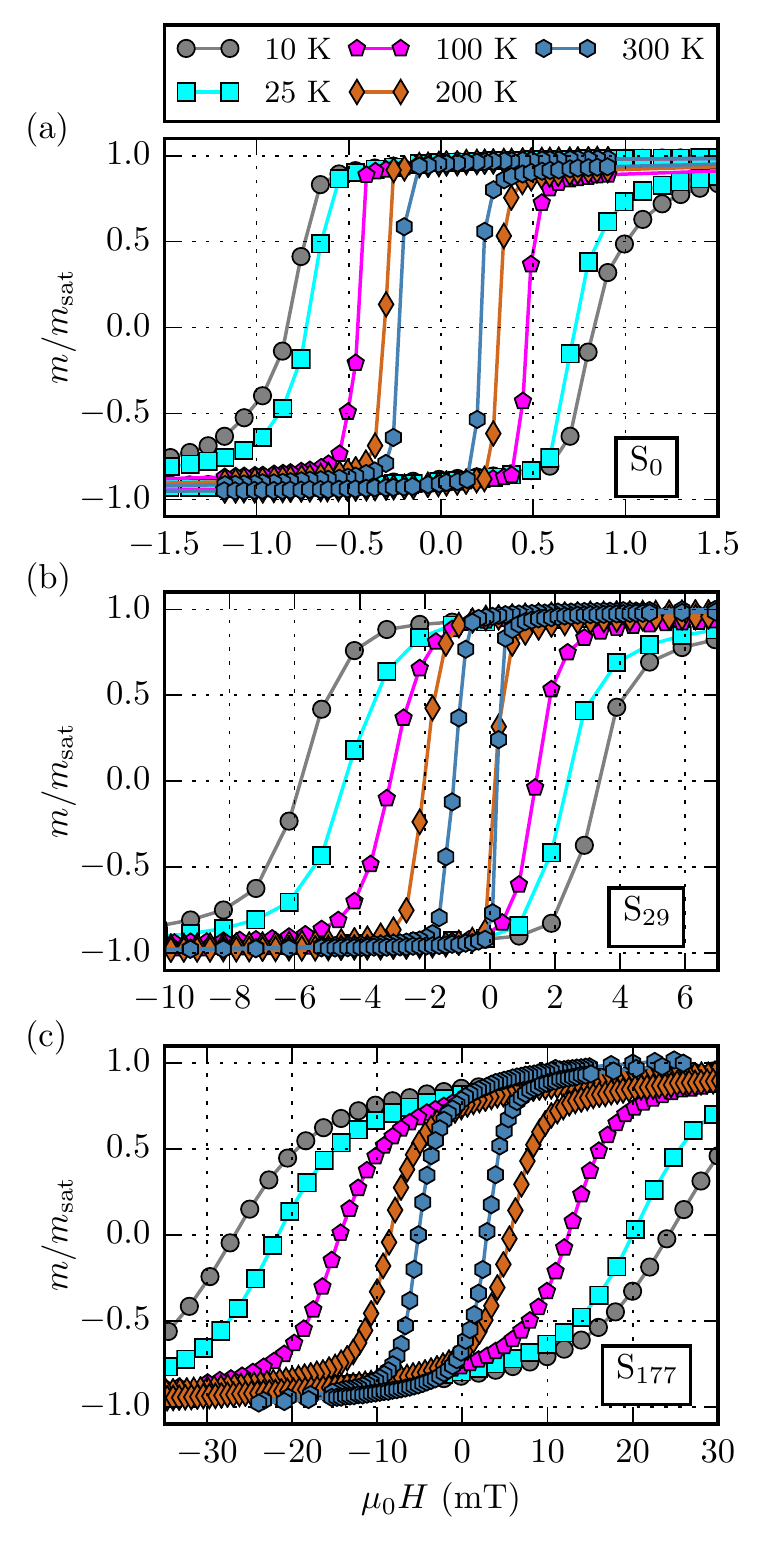}
        \caption{
            Hysteresis cycles at different temperatures indicated, following the FC protocol for (a) \Sa, (b) \Sb  and (c) \Sc samples.
            Note the different $x$-axis scales.
        }
        \label{fig:3}
    \end{figure}

    In Fig.~\ref{fig:3} the M-H measurements using the FC protocol are shown for (a) \Sa, (b) \Sb and (c) \Sc.
    From figures \ref{fig:3}(b) and \ref{fig:3}(c) it is seen that for samples \Sb and \Sc, the M-H loops are shifted along the field axis.
    In addition, coercive field enhancements are observed for \Sb and \Sc compared to the values found for \Sa at the same temperature.
    The presence of the exchange bias and the coercive enhancement are characteristics of EB systems.%
    \cite{Nogues-2005-ID171}
    It should be noted that the significant roughness observed in \Sb and \Sc could also contribute to the coercivity enhancement.%
    \cite{Li-1998-ID677}

    \begin{figure}
        \includegraphics{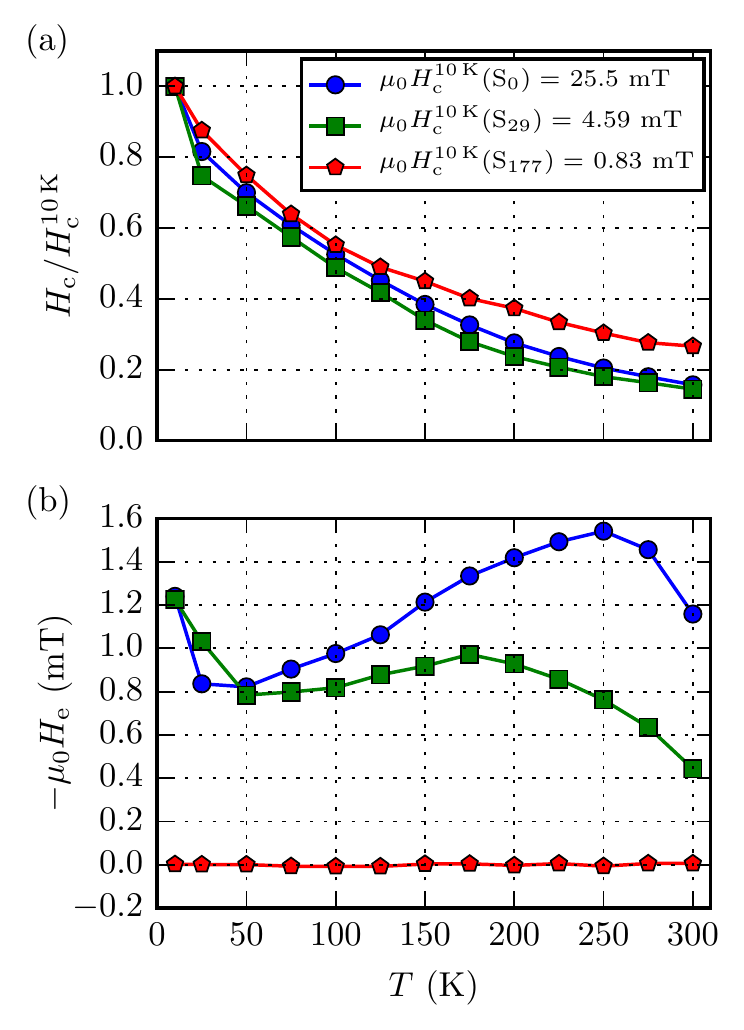}
        \caption{
            Temperature evolution following the FC protocol of (a) the reduced coercive field \( \Hc / \Hc^\mathrm{\SI{10}{K}} \)  and (b) the exchange field \He, for \Sa (red pentagon), \Sb (green squares) and \Sc (blue circles) samples.
            The \( \mu_0\Hc \) values at 10~K for the three samples are indicated in the legend of figure (a).
        }
        \label{fig:4}
    \end{figure}

    The behavior of \( \Hc(T) \) and \( \He(T) \) are extracted from the FC protocol M-H measurements for all the samples and are shown in Fig.~\ref{fig:4}.
    The coercive field \Hc decreases monotonically with increasing temperature for all the samples, as is typically expected in F materials and EB systems.
    This behavior therefore corresponds to previously reported experimental studies on polycrystalline and epitaxial BFO/F exchange biased systems.%
    \cite{Rao-2013-ID331, Naganuma-2011-ID127, He-2015-ID326, Martin-2008-ID112, Xue-2013-ID83}

    In Fig.~\ref{fig:4}(b) the \( \He(T) \) behavior of the various samples are shown.
    It is clear from the figure that \He is zero and does not vary with temperature for the unbiased \Sa sample.
    The \( \He(T) \) behavior of the \Sb and \Sc samples does not show a  monotonic decrease, but exhibits a sharp decrease at low temperature, displaying a peak at an intermediate temperature.
    In the present study this peak appears at 175~K for \Sb, and at 250~K for \Sc, respectively.
    This non-monotonic \( \He(T) \) behavior is a common behavior for BFO bilayers, in the sense that it is relevant to different F coupled to BFO.
    \cite{Xue-2013-ID83, Naganuma-2011-ID127, He-2015-ID326, Martin-2008-ID112}
    This behavior is also common in both polycrystalline\cite{Xue-2013-ID83}
    and epitaxial BFO.%
    \cite{Naganuma-2011-ID127, He-2015-ID326, Martin-2008-ID112}
    Thus, the driving mechanism for a non-monotonic \( \He(T) \) behavior of BFO/F should not depend on the BFO crystallographic arrangement, nor \tbfo or the thickness and composition of the F layer.

\subsection{Temperature dependent magnetization reversal using the Soeya protocol}

    As the thermal evolution of \He depends on thermal evolutions of the AF entities which are pinned, it is of interest to probe the BFO thermal activation energies present in our samples.
    In order to investigate the thermal behavior, the Soeya protocol was used.
    It may be noted that while the Soeya protocol has been performed on epitaxial exchange coupled BFO,%
    \cite{Safeer-2012-ID86}
    it has, however, never before been performed on polycrystalline exchange coupled BFO.

\begin{figure}
        \includegraphics{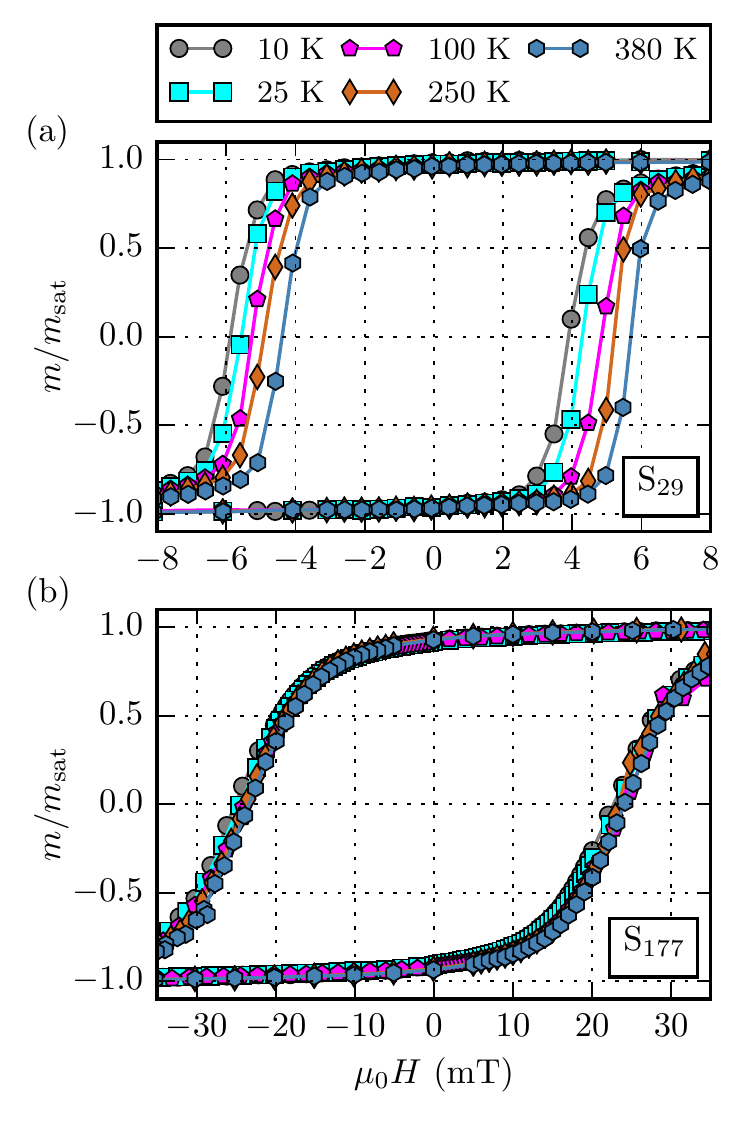}
        \caption{
            Hystereris cycles at different temperatures using the Soeya protocol for (a) \Sb, and (b) \Sc samples.
        }
        \label{fig:5}
    \end{figure}

\begin{figure}
        \includegraphics{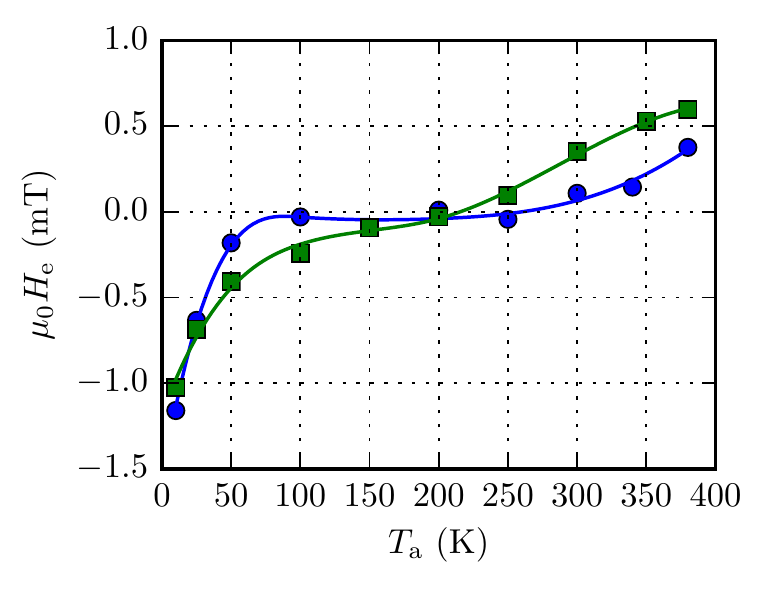}
        \caption{
            Evolution of \He using the Soeya protocol for \Sb (green squares) and \Sc (blue circles) samples.
        }
        \label{fig:6}
    \end{figure}

    The \Sb and \Sc hysteresis loops obtained from the Soeya protocol in the temperature range of 10~K to 380~K are shown in Fig.~\ref{fig:5}.
    The \( \He(\Ta) \) and \( \Hc(\Ta) \) curves were extracted from these hysteresis loops and are reported in Fig.~\ref{fig:6}.
    For both samples, \( \Hc(\Ta) \) is constant (not shown), as all the experiments were performed at the same \Tm.
    However, Fig.~\ref{fig:6} indicates that \He evolves with the activation temperature \Ta.
    \( \He(\Ta) \)  of both \Sb and \Sc exhibit similar two-step like behavior: i) the first step is seen before 100~K, ii) while second step is seen above 250~K.
	For both steps, \He presents a significant  variation with \Ta.
	In between these two steps, \( \He(\Ta) \) for \Sb and \Sc are different: in \Sb, it exhibits a positive slope with increasing \T, whereas for \Sc, it is constant.

    A two-step \( \He(\Ta) \) evolution in the Soeya protocol was previously reported for BFO/CoFeB epitaxial system.%
    \cite{Safeer-2012-ID86}
    This two-step reversal was attributed to two different contributions:
    at low temperature, the AF/F disordered interfacial spins would exhibit a spin-glass like behavior and would then be responsible for the first step of \( \He(\Ta) \), whereas domain wall depinning energy would be the driving mechanism for the second step.
    The behavior observed in Fig.~\ref{fig:6} are similar than those observed for epitaxially grown BFO/CoFeB, where \tbfo were chosen to be in the same thickness intervals.%
    \cite{Safeer-2012-ID86}
    The only difference between the present values of \He and those observed in epitaxially grown BFO is the magnitude of \He.
    This is expected as the \He magnitude  is related to the F thickness and magnetization, which were different in these studies.
    Thus, this common evolution of \He with \Ta indicates a driving mechanism independent of the long-range crystalline arrangement (i.e., epitaxial or polycrystalline arrangement) and of the nature of the F layer.
    Indeed, it is rather unlikely that two distinct mechanisms, the first one present in polycrystalline films and the second one present in epitaxial ones, would result in exactly the same \( \He(\Ta) \) behavior.
    This is even more unlikely for two different \tbfo intervals.
    If truly present, the driving mechanism proposed for the epitaxial BFO, involving spin-glass like interfacial disorder and domain wall depinning energy, should be valid for the polycrystalline samples.
    However, it is not expected that the domain wall depinning energy would be the same in polycrystalline and epitaxial systems:
    such an energy would certainly depend on the crystalline arrangement.
    Consequently, it is of interest to analyze this common \( \He(\Ta) \) two-step behavior in BFO, and the common non-monotonic \( \He(T) \) behavior considering that both phenomena are driven by a BFO physical property independent of its long-range crystalline arrangement.

    An inherent BFO property that can be considered is the canting of the BFO spins.
    This canting is present in either polycrystalline BFO or epitaxial BFO.%
	\cite{Ederer-2005-ID139, Ederer-2008-ID606, Naganuma-2007-ID626, Troyanchuk-2009-ID628, Ramazanoglu-2011-ID629, Marchand-2016-ID630}
    Such a canting results in a non-zero component of the BFO magnetic moment, oriented in a perpendicular direction compared to the non-canted case.
    Consequently, in a BFO/F bilayer, the exchange coupling energy resulting from this non-zero component will be a minimum in the perpendicular direction.
    This phenomenon is similar to the one proposed by Slonczewski to describe perpendicular exchange coupling in Fe/Cr multilayers,%
    \cite{Slonczewski-1991-ID542}
    involving a biquadratic exchange energy term.
    Furthermore, previous micromagnetic calculations confirmed that perpendicular coupling does result when canting is allowed.%
	\cite{Koon-1997-ID152}

    In the BFO/F systems discussed here, the biquadratic coupling promoted by intrinsic BFO properties such as the canted spins should contribute to the exchange bias properties.
    Indeed, it was recently shown by simulation that the presence of biquadratic coupling in AF/F systems results in a non-monotonic behavior of \( \He(T) \), with the presence of a peak at intermediate temperature.%
    \cite{Hu-2004-ID120}
    This supports the idea of a driving mechanism relying on an inherent BFO property, that is, the presence of canted spins being at the origin of the common temperature dependent phenomenon observed in epitaxial and polycrystalline BFO/F systems.
    To probe the presence of a biquadratic coupling in the samples studied here, the angular dependence of the magnetization reversal was measured using VSM measurements at RT and at \SI{77}{K}.

\subsection{Temperature dependent magnetization reversal using azimuthal measurements}

\begin{figure}
        \includegraphics{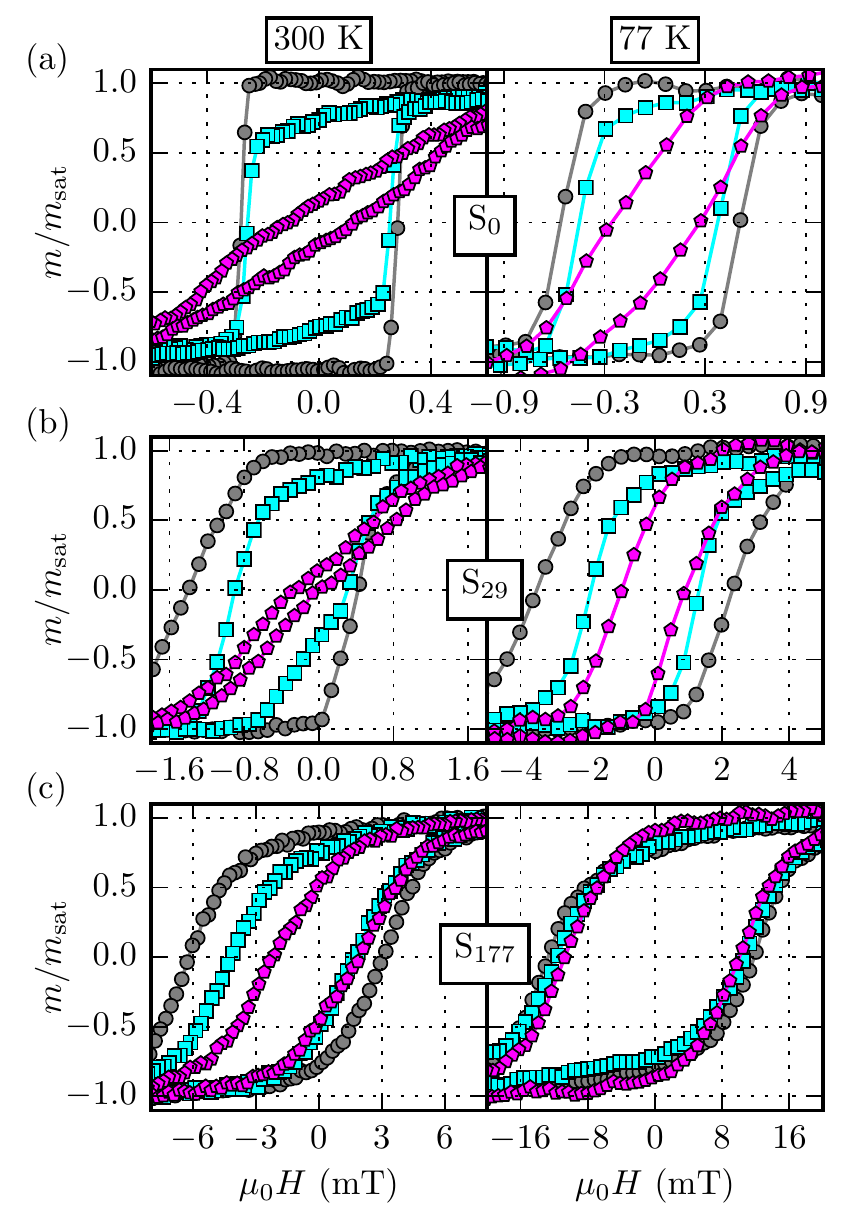}
        \caption{
            Magnetization versus field curves obtained using the VSM for \( T = \SI{300}{K} \) (left) and \( T = \SI{77}{K} \) (right), for various thicknesses \tbfo (a) 0~nm, (b) 29~nm and (c) 177~nm.
            Measurements were performed with field angles at \SI{0}{\degree} (gray circles), \SI{45}{\degree} (cyan squares) and \SI{90}{\degree} (magenta pentagons) away from the uniaxial easy anisotropy axis (i.e \( \varphi \approx \SI{20}{\degree}\) for \Sa, \( \varphi \approx \SI{10}{\degree}\) for \Sb and \( \varphi = \SI{0}{\degree}\) for \Sc).
        }
        \label{fig:7}
    \end{figure}

\begin{figure}
        \includegraphics{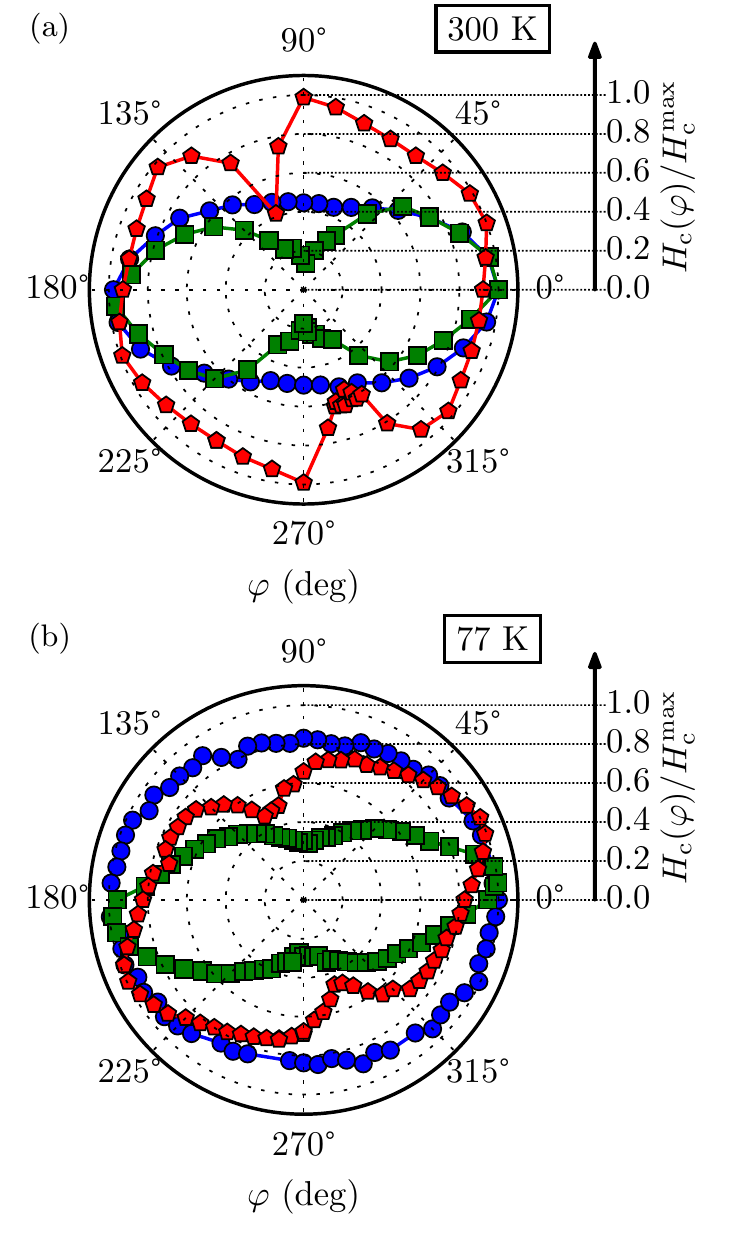}
        \caption{
            Azimuthal evolution of the reduced coercive field \( \Hc(\varphi)/\Hc^\mathrm{max} \) at (a) \SI{300}{K} and (b) \SI{77}{K}, with \( t_\mathrm{BFO} = \) \SI{0}{nm} (red pentagons), \SI{29}{nm} (green squares) and \SI{177}{nm} (blue circles).
        }
        \label{fig:8}
    \end{figure}

\begin{figure*}
        \includegraphics{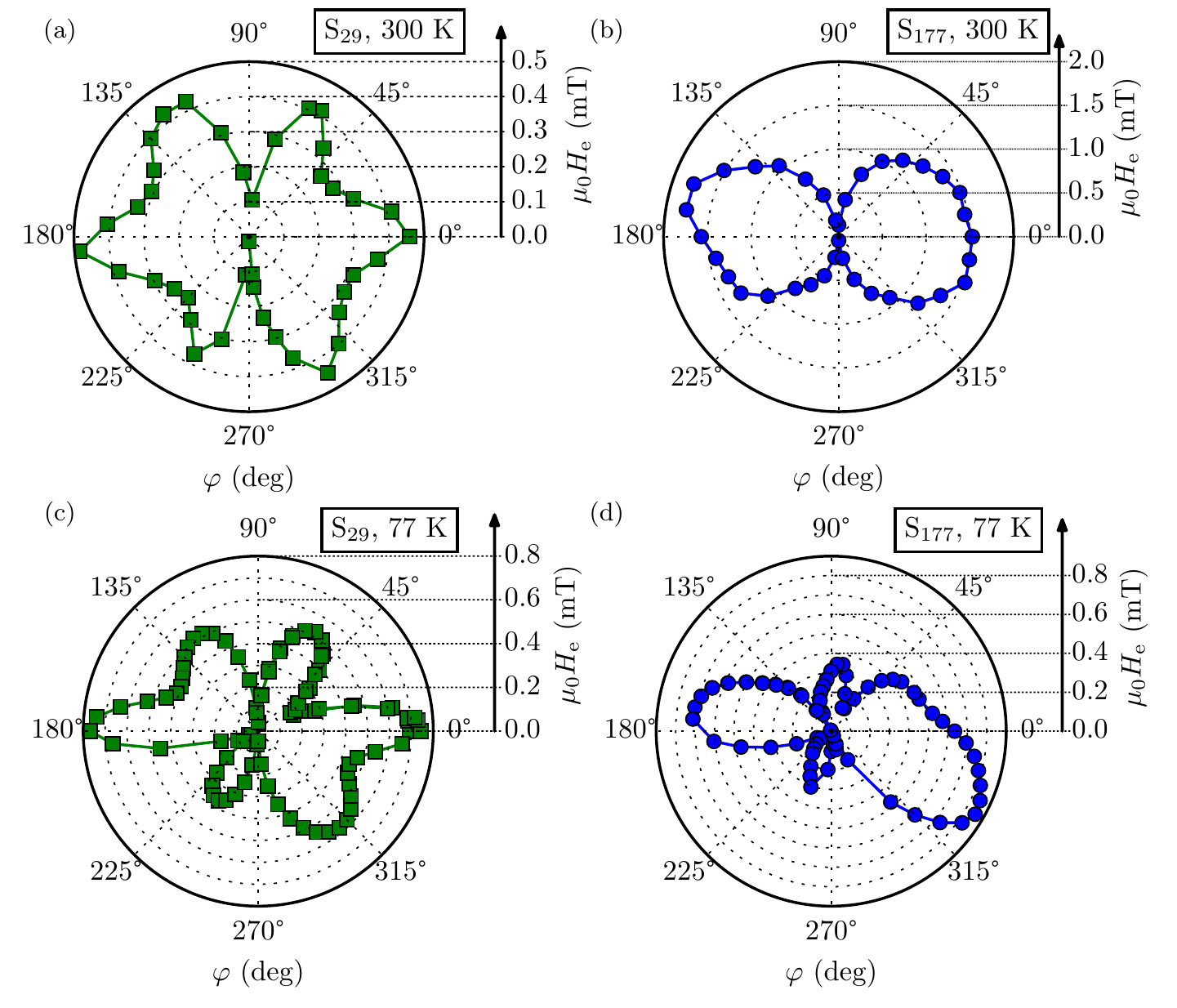}
        \caption{
            Angular evolution of \He: (a) and (b), at \SI{300}{K}; (c) and (d), at \SI{77}{K}.
            The samples measured are \Sb (green squares) and \Sc (blue circles).
        }
        \label{fig:9}
    \end{figure*}

    Magnetization reversal loops were measured at \SI{77}{K} and RT, applying the external field \(H\) at various \( \varphi \) angles.
    Results are shown in Fig.~\ref{fig:7}.
    \( \Hc(\varphi) \) and \( \He(\varphi) \) obtained from the measurements in Fig.~\ref{fig:7} are shown in Fig.~\ref{fig:8} and Fig.~\ref{fig:9}, respectively.
    For all samples and at both temperatures, M-H behaviors are shown to be strongly dependent on the thickness of BFO.

    In Fig.~\ref{fig:8} the \Hc angular dependence is shown for the sample in which BFO was absent (\Sa), for both measurements at RT and 77~K.
    \( \Hc(\varphi) \) exhibits a maximum at \( \varphi = \SI{20}{\degree} \) and a minimum at \( \varphi = \SI{110}{\degree} \) shown in Fig.~\ref{fig:8}(a) at 300~K.
    This confirms the uniaxial character of the non-coupled \py layer anisotropy.
    The reversal cycles at \( \varphi = \SI{20}{\degree} \) and \( \varphi = \SI{110}{\degree} \) are typical of an uniaxial easy axis loop for \( \varphi = \SI{20}{\degree} \) and hard axis for \( \varphi = \SI{110}{\degree} \), as shown in Fig.~\ref{fig:7}(a).
    The hysteresis observed along the hard axis implies an angular dispersion of the easy axis.
    Thus, Py is uniaxial with a \SI{20}{\degree} misaligned easy axis relatively to \Hdep.
    This analysis is valid for both temperatures.

    The Py layer coupled with a thin BFO layer in the \Sb sample exhibits a coercive enhancement (at 300~K) compared to \Sa, and an angular dependent shift of the hysteresis loop along the field axis as shown in Fig.~\ref{fig:8}(a) and in Fig.~\ref{fig:9}(a), respectively.
    The \Hc angular dependence exhibit a maximum at \( \varphi = \SI{5}{\degree}\) and a minimum at \( \varphi = \SI{95}{\degree}\) as shown in Fig.~\ref{fig:8}(a).
    At 300~K, a two-step magnetization reversal process is observed when \(H\) is at \SI{95}{\degree} (i.e., along the \Hc minimum).
    Such a two-step reversal reveals a minimum of the magnetic energy along that direction, as expected from the contribution of a biquadratic coupling term which favors a perpendicular orientation of the F moments relatively to the AF ones.%
    \cite{Moran-1998-ID618, Dumm-2000-ID40, Michel-1998-ID543, McCord-2009-ID41, McCord-2008-ID59, Dekadjevi-2011-ID97}
    It reveals that the canting in BFO plays a key role for the \Sb magnetization reversal and support the hypothesis of the canting being a driving mechanism for the exchange biased properties and their temperature dependence.
    Furthermore, it should be noted that the two-step magnetization reversal reported here was also observed in a previous experimental study on epitaxial \( \mathrm{Co_{75}Fe_{25}/BFO} \) and \( \mathrm{Co_{50}Fe_{50}/BFO} \).%
    \cite{Naganuma-2011-ID127}
    However, this feature was not discussed by the authors.
    In this previous work, \( \He(T) \) exhibited a similar non-monotonic behavior to the one reported here in Fig.~\ref{fig:4}(b).

    It should be noted that the exchange coupling of the Py with the thin BFO layer does not modify the overall shape of the \( \Hc(\Sb) \) angular dependence (relatively to the uncoupled Py in \Sa), despite the biquadratic contribution.
    The absence of a fourfold symmetry arising from a biquadratic contribution suggests that the uniaxial anisotropy energy is greater than the biquadratic contribution to the magnetic anisotropy energy.
    Thus, the evidence for a contribution which favors a \SI{90}{\degree} phase is the two-step magnetization reversal process along the perpendicular direction to the uniaxial easy axis,%
    \cite{Dumm-2000-ID40, Michel-1998-ID543, McCord-2009-ID41, Dekadjevi-2011-ID97}
    as previously discussed in Fig.~\ref{fig:7}(b).
    This two-step magnetization reversal and the \Hc angular dependence demonstrate that the uniaxial anisotropy dominates the biquadratic contribution in \Sb.
    At \SI{77}{K}, the overall shape of the \Hc angular dependence is similar but the minimum observed along the uniaxial hard axis is less pronounced than the one observed at \SI{300}{K} as shown in Fig.~\ref{fig:8}(b).
    It indicates that the anisotropy dispersion is more pronounced at \SI{77}{K} than at \SI{300}{K}.
    This is confirmed by the large opening of the hysteresis curves shown in Fig.~\ref{fig:7}.

    The Py layer coupled with a thick BFO layer in the \Sc sample exhibits an enhanced coercivity relatively to \Sa and \Sb, as shown in Fig.~\ref{fig:7}(c).
    In Fig.~\ref{fig:8}(a), the \Hc angular dependence of \Sc at \SI{300}{K} corresponds to an ellipse.
    There is no local minimum at \SI{90}{\degree} of the easy axis, indicating a large dispersion of the anisotropy axis.
    The angular dependence of \Hc at \SI{77}{K} is quasi-circular revealing a random anisotropy dispersion.

    The \He angular dependence in exchange biased systems depends on the ratio of the unidirectional and anisotropic energy contributions.%
    \cite{Spenato-2007-ID44, Hu-2002-ID609, Jimenez-2009-ID385, *Jimenez-2011-ID386}
    At RT,  the \He angular dependence for \Sb is characteristic of a misaligned configuration of the anisotropy axes.
    Indeed, the presence of a star-like azimutal shape is well-known and can be reproduced using a coherent rotation model.%
    \cite{Spenato-2007-ID44, Jimenez-2009-ID385, *Jimenez-2011-ID386}
    In such a shape, the misalignement is revealed by the assymetry of the arms.
    Thus, as shown for \Sb in Fig.~\ref{fig:9}(a), the misalignment is indicated by the reduced \He maximum value at \( \varphi = \SI{65}{\degree} \)  and \( \varphi = \SI{245}{\degree}\)) relatively to the ones at \( \varphi = \SI{120}{\degree} \) and \(\varphi = \SI{300}{\degree}\), respectively.
    At RT in Fig.\ref{fig:9}(b), the \( \He(\Sc) \) angular dependence exhibits two asymmetric lobes, relatively to the easy axis.
    In a recent work, the presence of two asymmetric exchange lobes in BFO/Py could be reproduced using a coherent rotation model considering a biquadratic-like anisotropy and a small \SI{5}{\degree} misalignment between the anisotropy axis directions.%
    \cite{Hauguel-2011-ID4}
    For both samples, the \He angular shape is strongly temperature dependent since the curves obtained at \SI{77}{K} are much different than the ones obtained at \SI{300}{K}, as shown in Fig.~\ref{fig:9}.
    Since the \He angular dependences are strongly dependent of the ratio between effective anisotropy constants, this temperature dependence is expected as \( \He(T) \) and \( \Hc(T) \) evolves in a much different manner with temperature as shown in Fig.~\ref{fig:9} and Fig.~\ref{fig:8}, indicating a much different evolution of the various effective anisotropies in a given sample.
    The thermal dependent azimutal measurements demonstrate complex arrangements of the anisotropy axis and are in agreement with the presence of a biquadratic contribution to the magnetic energy of the BFO/Py studied here.
    A biquadratic driving mechanism for the thermal properties of BFO/F systems induced by the canting of the BFO spins depends neither on the long-range crystalline arrangement of the BFO nor on the F layer, as it is an intrinsic property of BFO.
    It is in agreement with previously reported \( \He(T) \) and \( \Hc(T) \) behaviors following a FC protocol and the \( \He(\Ta) \) behavior following the Soeya protocol, in polycrystalline and epitaxial BFO.

\section{Conclusion}

    In the current contribution, the thermal dependences of exchange bias properties are probed for three different BFO thicknesses (0~nm, 29~nm and 177~nm).
    These were chosen as they represent three regions of interest in the magnetic behavior of the BFO/Py system:
    i) \Sa corresponds to an unbiased sample;
    ii) \( \tbfo = \SI{29}{nm} \) (\Sb) is just above \tc, an interval where \He(\tbfo) is strongly thickness dependent;
    and iii) \( \tbfo = \SI{177}{nm} \) (\Sc) is far larger than \Tc, an interval where \He(\tbfo) is thickness independent.
    Three different methods were employed to study the thermal dependence of the exchange bias of BFO/Py system.

    The first approach consists of a field cool procedure.
    It shows that \( \Hc(T) \) decreases monotically with increasing temperature for all BFO thicknesses, whereas \( \He (T) \) exhibits a non-monotonic behavior, with the presence of a middle temperature range peak, when the Py layer is exchange coupled with the BFO one.
    This \He and \Hc temperature behavior confirm previous experimental behaviors on epitaxial and polycrystalline BFO/F systems, demonstrating that this \( \He(T) \) non-monotonic behavior is independent of the BFO crystalline arrangement, thickness, and independent on the F nature.

    The second thermal approach was carried out on the exchange-coupled samples (i.e., \Sb and \Sc) and consists of the Soeya protocol which relates to the BFO thermal activation energies at the origin of the exchange bias properties.
    The evolution of \He with the activation temperature presents a two-step evolution for both samples.
    This behavior in the polycrystalline BFO/Py system studied here is identical to the one observed in epitaxial BFO.
    Consequently, the thermal behavior of the BFO/F exchange bias field probed here is shown to be independent of the crystalline arrangement, thickness, and independent on the F nature.
    It indicates that the driving mechanism for a non-monotonic \He in exchange coupled BFO systems relies on a physical property or properties not related to the ones discussed above.
    An intrinsic driving property of BFO is proposed as being this driving mechanism: the canting of the BFO spins leading to a biquadratic contribution to the exchange coupling.

    The third thermal approach was to probe the magnetization reversal angular dependencies at RT and at \SI{77}{K}, as it provides information concerning axial and unidirectional properties.
    For sample \Sb, the magnetization reversal angular dependence demonstrates the presence of a biquadratic contribution.
    For all samples, the temperature dependence of the angular behavior of the magnetization reversal agrees with the presence of a biquadratic contribution and is driven by the anisotropic ratio, including the presence of misalignments.

    Therefore, a common mechanism of a biquadratic contribution, for driving temperature dependent exchange bias properties, is supported by the thermal dependent studies presented here.
    It is of interest to implement explicitly such a mechanism in theoretical approaches in order to predict and tailor the thermal dependent exchange bias properties in BFO systems.

\begin{acknowledgments}
    We acknowledge the french microscopy network METSA for TEM experiments, and the PIMM-DRX shared facility of the university of Brest for XRD measurements.
    The authors also thank R. Tweed for reading through the manuscript.
    A. R. E. Prinsloo and C.J. Sheppard thanks the FRC/URC of UJ and SA-NRF for the  financial support (Protea funding: 85059).
    A. M. Strydom thanks the FRC/URC of UJ and the SA-NRF (93549) for financial assistance.
    D. T. Dekadjevi thanks the MAEDI and MENESR for the  financial support (PHC PROTEA 2014 N\textsuperscript{\b{o}}29785ZA9).

    All the figures were generated using the Matplotlib software.\cite{Hunter-2007-ID454}
\end{acknowledgments}

\bibliography{library_extract}

\end{document}